\documentstyle[12pt]{article}
\def\thebibliography#1{\bigskip\section*{\centering
References\\}\bigskip\list
{\arabic{enumi}.}{\settowidth\labelwidth{#1}\leftmargin\labelwidth
\advance\leftmargin\labelsep
\usecounter{enumi}}
\def\newblock{\hskip .11em plus .33em minus .07em}
\sloppy\clubpenalty4000\widowpenalty4000
\sfcode`\.=1000\relax}

\def\op#1{\mathop{\fam0 #1}\limits}

\newcommand{\ben}{\begin{eqnarray}}
\newcommand{\een}{\end{eqnarray}}

\newcommand{\be}{\begin{eqnarray*}}
\newcommand{\ee}{\end{eqnarray*}}
\newcommand{\bea}{\begin{eqalph}}
\newcommand{\eea}{\end{eqalph}}
\newcommand{\cL}{{\cal L}}

\newcommand{\R}{{\bf R}}
\newcommand{\C}{{\bf C}}

\newcommand{\cD}{{\cal D}}

\newcommand{\dl}{\delta}
\newcommand{\la}{\lambda}

\newcommand{\om}{\omega}
\newcommand{\ot}{\otimes}

\newcommand{\m}{\mu}
\newcommand{\n}{\nu}
\newcommand{\g}{\gamma}
\newcommand{\G}{\Gamma}
\newcommand{\e}{\epsilon}

\newcommand{\Si}{\Sigma}
\newcommand{\si}{\sigma}

\newcommand{\w}{\wedge}

\newcommand{\wt}{\widetilde}
\newcommand{\wh}{\widehat}

\newcommand{\dr}{\partial}

\newcounter{eqalph}
\newcounter{equationa}

\newenvironment{eqalph}{\stepcounter{equation}
\setcounter{equationa}{\value{equation}}
\setcounter{equation}{0}

\begin{eqnarray}}{\end{eqnarray}
\setcounter{equation}{\value{equationa}}}

\begin{document}
\hbox{}

\centerline{\large\bf Dirac Equation in Gauge and Affine-Metric}
\medskip

\centerline{\large\bf Gravitation Theories}
\bigskip

\centerline{\sc Giovanni Giachetta}

\medskip

\centerline{Department of Mathematics and Physics}

\centerline{University of Camerino, 62032 Camerino, Italy}

\centerline{E-mail: mangiarotti@camvax.unicam.it}
\medskip

\centerline{\sc Gennadi Sardanashvily}
\medskip

\centerline{Department of Theoretical Physics}

\centerline{Moscow State University, 117234 Moscow, Russia}

\centerline{E-mail: sard@grav.phys.msu.su}
\bigskip

\begin{abstract}
We show that the covariant derivative of Dirac fermion fields in the presence
of a general linear connection on a world manifold is universal for Einstein's,
gauge and affine-metric gravitation theories.
\end{abstract}

\section{}

We follow the gauge approach to the description of gravitational interactions.

In reality, one observes three types of fields: gravitational
fields,  fermion fields and gauge
fields associated with internal symmetries. We are not concerned here with
Higgs fields whose dynamics remains elusive.

If the gauge invariance under internal symmetries is kept in the presence of
a gravitational field, Lagrangian densities of gauge fields must depend on a
metric gravitational field only.

Different spinor models of fermion matter have been suggested. At present, all
observable fermion particles are Dirac fermions. Therefore, we restrict our
consideration to Dirac fermion fields in gravitation theory.

In the gauge gravitation theory, gravity is represented by pairs $(h,A_h)$
of gravitational fields $h$ and associated Lorentz connections
$A_h$ \cite{heh,sard92}. The connection $A_h$ is usually identified with
both a connection on a world manifold $X$ and a spinor connection on the
the spinor bundle $S_h\to X$ whose sections describe Dirac fermion fields
$\psi_h$ in the presence of the gravitational field $h$. The problem arises
when Dirac fermion fields are described in the framework of the affine-metric
gravitation theory. In this case, the fact that a world connection is
some Lorentz connection may result from the field
equations, but it can not be assumed in advance. There are models where the
world connection is not a Lorentz connection \cite{heh}. Moreover, it may
happen that a world connection is the Lorentz connection with respect
to different gravitational fields \cite{art}. At the same time, a
Dirac fermion field can be regarded only in a pair $(h,\psi_h)$ with a certain
gravitational field $h$.

Indeed, one must define the representation
of cotangent vectors to $X$ by the Dirac's $\g$-matrices in order to
construct the Dirac operator.
Given a tetrad gravitational field $h(x)$, we have the representation
$$
\g_h: dx^\mu \mapsto  \wh dx^\mu =h^\mu_a\g^a.
$$
However, different gravitational fields $h$ and $h'$ yield the
nonequivalent representations $\g_h$ and $\g_{h'}$.

It follows that, fermion-gravitation pairs
$(h,\psi_h)$ are described by sections of the composite spinor bundle
\begin{equation}
S\to\Si\to X \label{L1}
\end{equation}
where  $\Si\to X$  is the bundle of gravitational fields $h$
where values of $h$ play the role of
parameter coordinates, besides the familiar world coordinates.
\cite{sard92,sard1}. In particular, every spinor bundle $S_h\to X$ is
isomorphic to the restriction of $S\to\Si$ to $h(X)\subset \Si$. Performing
this restriction, we come to the familiar case of a field model in the
presence of a gravitational field $h(x)$. The feature of the dynamics of field
systems on the composite bundle (\ref{L1}) lies in the fact that we
have the modified covariant differential of fermion fields which depend on
derivatives of gravitational fields $h$.

As a consequence,
we  get the following covariant
derivative of Dirac fermion fields in the presence of a gravitational field
$h(x)$:
\begin{equation}
\wt D_\la  =\dr_\la -\frac12A^{ab}{}_\m^c (\dr_\la h^\m_c +
K^\m{}_{\nu\la}  h^\nu_c)I_{ab}, \label{K101}
\end{equation}
$$
A^{ab}{}_\m^c =\frac12(\eta^{ca}h^b_\m -\eta^{cb}h^a_\m),
$$
where $K$ is a general linear connection on a world manifold $X$, $\eta$
is the Minkowski metric, and
$$
I_{ab}=\frac14[\g_a,\g_b]
$$
are generators of the spinor group $L_s= SL(2,\C)$.

Let us emphasize that
the connection
\begin{equation}
\wt K^{ab}{}_\la = A^{ab}{}_\m^c (\dr_\la h^\m_c +
K^\m{}_{\nu\la}  h^\nu_c) \label{K102}
\end{equation}
is not the connection
$$
K^k{}_{m\la}=h^k_\m(\dr_\la h^\m_m +K^\m{}_{\nu\la} h^\nu_m) = K^{ab}{}_\la
(\eta_{am}\dl^k_b -\eta_{bm}\dl^k_a)
$$
written with respect to the reference frame $h^a=h^a_\la dx^\la$, but there is
the relation
\begin{equation}
\wt K^{ab}{}_\la=\frac12(K^{ab}{}_\la -K^{ba}{}_\la).\label{K50}
\end{equation}
If $K$ is a Lorentz connection $A_h$, then the connection $\wt K$
(\ref{K102}) consists with $K$ itself.

The covariant derivative (\ref{K101}) has been considered by several authors
\cite{ar,pon,tuc}. The relation (\ref{K50}) correspond to the canonical
decomposition of the Lie algebra of the general linear group. By the
well-known theorem \cite{kob}, every general linear connection being
projected onto the Lie algebra of the Lorentz group yields a Lorentz
connection.

In our opinion, the advantage of the covariant derivative (\ref{K101}),
consists in the fact that, being derived in the framework of the gauge
gravitation theory, it may be also applied to the affine-metric
gravitation theory and the conventional Einstein's gravitation theory.
We are not concerned here with the general
problem of equivalence of metric, affine and affine-metric theories of
gravity \cite{fer,mag}.
At the same time, when $K$ is
the Levi-Civita connection of $h$, the Lagrangian density of fermion
fields which utilizes the covariant derivative (\ref{K101}) comes to that
in the Einstein's gravitation theory. It follows that the configuration
space of metric (or tetrad) gravitational fields and general linear
connections  may play the role of the universal configuration space of
realistic gravitational models. In particular, one then can think of the
generalized Komar superpotential as being the universal superpotential
of energy-momentum of gravity  \cite{cam2}.

\section{}

We follow the geometric approach to field theory when classical fields are
described by global sections of a bundle $Y\to X$ over a world manifold $X$.
Their dynamics is phrased in terms of jet manifolds
\cite{sard,sard0,sau}.

As a shorthand, one can say that the
$k$-order jet manifold $J^kY$ of a bundle $Y\to X$
comprises the equivalence classes
$j^k_xs$, $x\in X$, of sections $s$ of $Y$ identified by the first $k+1$
terms of their Taylor series at a point $x$.  Recall that
a $k$-order differential operator on sections of a bundle $Y\to X$ is defined
to be a bundle morphism of the bundle $J^kY\to X$ to a vector
bundle over $X$.

In particular, given bundle
coordinates $(x^\m,y^i)$ of a bundle $Y$, the first order jet manifold $J^1Y$
of $Y$ is endowed with the coordinates $ (x^\m,y^i,y^i_\m)$ where
$$
y^i_\m(j^1_x s) =\dr_\m s^i(x).
$$

There is the 1:1
correspondence between the connections on the bundle $Y\to X$ and the global
sections
$$
\G =dx^\la\ot(\dr_\la+\G^i_\la\dr_i)
$$
of the affine jet bundle $J^1Y\to Y$. Every connection $\G$ on $Y\to X$ yields
the first order differential operator
\be
&& D_\G:J^1Y\op\to_YT^*X\op\ot_YVY,\\
&&D_\G=(y^i_\la-\G^i_\la)dx^\la\ot\dr_i,
 \ee
on $Y$ which is called the covariant differential relative to the
connection $\G$.
We denote by $VY$ the vertical tangent bundle of $Y$.

In the first order Lagrangian formalism,
the first order
jet manifold $J^1Y$ of $Y$ plays the role of
the finite-dimensional
configuration space
of fields represented by sections $s$ of a bundle $Y\to X$.
A first order
Lagrangian density is defined to be an exterior horizontal density
\be
&& L: J^1Y\to\op\w^nT^*X, \qquad n=\dim X,\\
&&L=\cL(x^\m,y^i,y^i_\m)\om, \qquad \om=dx^1\w ...\w dx^n,
\ee
on the bundle $J^1Y\to X$.

Note that, since the jet bundle $J^1Y\to Y$ is affine, every polynomial
Lagrangian density of field theory factors through
$$
L:J^1Y\op\to^D T^*X\op\ot_YVY\to\op\w^nT^*X
$$
where $D$ is the covariant differential relative to some connection on $Y$.

\section{}

Let us consider the gauge theory of gravity and fermion fields.
By $X$ is further meant an oriented 4-dimensional world manifold which
satisfies the well-known topological conditions in order that
gravitational fields and spinor structure can exist on $X$. To
summarize these conditions, we assume that $X$ is not compact and
that the tangent bundle of $X$ is trivial.

Let $LX$ be the principal bundle of oriented linear frames in tangent spaces to
$X$. In gravitation theory, its structure group
$GL^+(4,{\bf R})$
is reduced to the connected Lorentz group
$ L=SO(1,3).$
It means that there exists a reduced subbundle $L^hX$ of $LX$ whose
structure group is $L$.
In accordance with the well-known theorem, there is
the 1:1 correspondence between the reduced $L$ subbundles $L^hX$ of
$LX$ and the global
sections $h$ of the quotient bundle
\begin{equation}
\Si:=LX/L\to X. \label{5.15}
\end{equation}
These sections $h$ describe gravitational fields on $X$, for the bundle
(\ref{5.15}) is the 2-folder covering of the bundle of
pseudo-Riemannian metrics on $X$.

Given a section $h$ of $\Si$, let $\Psi^h$ be an atlas of $LX$ such that the
corresponding local sections $z_\xi^h$ of $LX$ take their values into $L^hX$.
With respect to $\Psi^h$ and a
holonomic atlas $\Psi^T=\{\psi_\xi^T\}$ of $LX$, a gravitational field $h$
can be represented by a family of $GL_4$-valued tetrad functions
\begin{equation}
h_\xi=\psi^T_\xi\circ z^h_\xi,\qquad
dx^\la= h^\la_a(x)h^a. \label{L6}
\end{equation}

By the Lorentz connections $A_h$ associated with a gravitational field $h$
are meant the principal connections on the reduced subbundle $L^hX$ of $LX$.
They give rise to principal connections on $LX$ and to spinor connections on
the $L_s$-lift $P_h$ of $L^hX$.

There are different ways to introduce Dirac fermion fields. We follow the
algebraic approach.

Given a Minkowski space $M$, let
$ Cl_{1,3}$ be the complex Clifford algebra generated by elements
of $M$. A spinor space $V$ is defined to be a
minimal left ideal of $ Cl_{1,3}$  on
which this algebra acts on the left. We have the representation
$$
\g: M\ot V \to V
$$
of elements of the Minkowski space $M\subset Cl_{1,3}$ by
Dirac's matrices $\g$ on $V$.

 Let us consider a bundle of complex Clifford algebras $ Cl_{1,3}$ over $X$
whose structure group is the Clifford group of invertible elements of $
Cl_{1,3}$. Its subbundles are both a spinor bundle $S_M\to X$ and the bundle
$Y_M\to X$ of Minkowski spaces of generating elements of  $ Cl_{1,3}$.
To describe Dirac fermion fields on a world manifold $X$, one must
require $Y_M$ to be isomorphic to the cotangent bundle $T^*X$
of $X$. It takes place if there exists a reduced $L$ subbundle $L^hX$ such
that
\[Y_M=(L^hX\times M)/L.\]
Then, the spinor bundle
\begin{equation}
S_M=S_h=(P_h\times V)/L_s\label{510}
\end{equation}
is associated with the $L_s$-lift $P_h$ of $L^hX$. In this case, there exists
the representation
\begin{equation}
\g_h: T^*X\ot S_h=(P_h\times (M\ot V))/L_s\to (P_h\times
\g(M\times V))/L_s=S_h \label{L4}
\end{equation}
of cotangent vectors to a world manifold $X$ by Dirac's $\g$-matrices
on elements of the spinor bundle $S_h$. As a shorthand, one can write
\[\wh dx^\la=\g_h(dx^\la)=h^\la_a(x)\g^a.\]

Given the representation (\ref{L4}), we shall say that sections of
the spinor bundle $S_h$ describe Dirac fermion fields in the presence of
the gravitational field $h$. Indeed,
let
$$
A_h =dx^\la\ot (\dr_\la +\frac12A^{ab}{}_\la
I_{ab}{}^A{}_B\psi^B\dr_A)
$$
be a principal connection on $S_h$. Given
the corresponding covariant differential $D$ and the
representation $\g_h$ (\ref{L4}), one can construct the Dirac
operator \begin{equation}
\cD_h=\g_h\circ D: J^1S_h\to T^*X\op\ot_{S_h}VS_h\to VS_h, \label{I13}
\end{equation}
\[\dot y^A\circ\cD_h=h^\la_a\g^{aA}{}_B(y^B_\la-\frac12A^{ab}{}_\la
I_{ab}{}^A{}_By^B)\]
on the spinor bundle $S_h$.

Different  gravitational fields $h$ and $h'$ define nonequivalent
representations $\gamma_h$ and $\gamma_{h'}$.
It follows that a Dirac fermion field must be regarded only in a pair with
a certain gravitational field. There is the 1:1 correspondence
between these pairs and sections of the composite spinor bundle (\ref{L1}).

\section{}

By a composite bundle is meant the composition
\begin{equation}
Y\to \Si\to X. \label{1.34}
\end{equation}
of a bundle $Y\to X$ denoted by $Y_\Si$ and a bundle $\Si\to X$.
It is coordinatized by
$( x^\la ,\si^m,y^i)$
where $(x^\m,\si^m)$ are coordinates  of
$\Si$ and $y^i$ are the fiber coordinates of $Y_\Si$.
We further assume that $\Si$ has a global section.

The application of composite bundles to field theory is
founded on the following \cite{sard9}. Given
a global section $h$ of $\Sigma$, the restriction $ Y_h$
of $Y_\Si$ to $h(X)$ is a subbundle
of $Y\to X$. There is the 1:1 correspondence between
the global sections $s_h$ of $Y_h$ and the global sections of
the composite bundle (\ref{1.34}) which cover $h$.
Therefore, one can think of sections $s_h$ of $Y_h$ as
describing fermion fields in the presence of a background parameter
field $h$, whereas sections
of the composite bundle $Y$ describe all the pairs $(s_h,h)$.
The configuration space of these pairs is the
first order jet manifold $J^1Y$ of the composite bundle $Y$.

The feature of the dynamics of field systems on composite bundles consists in
the following.

Every connection
$$
A_\Si=dx^\la\ot(\dr_\la+\wt A^i_\la\dr_i)
+d\si^m\ot(\dr_m+A^i_m\dr_i)
$$
 on the bundle $Y_\Si$  yields
the horizontal splitting
$$
VY=VY_\Si\op\oplus_Y (Y\op\times_\Si V\Si),
$$
\[\dot y^i\dr_i + \dot\si^m\dr_m=
(\dot y^i -A^i_m\dot\si^m)\dr_i + \dot\si^m(\dr_m+A^i_m\dr_i).\]
Building on this splitting, one can construct
the first order differential operator
\ben
&&\wt D:J^1Y\to T^*X\op\ot_Y VY_\Si,\nonumber\\
&&\wt D=dx^\la\ot(y^i_\la-\wt A^i_\la -A^i_m\si^m_\la)\dr_i,\label{7.10}
\een
on the composite
bundle $Y$. This operator posesses the following property.

Given a global section $h$ of $\Si$, let $\G$ be a connection on $\Si$
whose integral section is $h$, that is, $\G\circ h = J^1h$.
It is readily observed that the differential
(\ref{7.10}) restricted to $J^1Y_h\subset J^1Y$ comes
to the familiar covariant
differential relative to the connection
$$
A_h=dx^\la\ot[\dr_\la+(A^i_m\dr_\la h^m +\wt A^i_\la)\dr_i]
$$
on $Y_h$.
Thus, it is $\wt D$ that
we may utilize in order to construct a Lagrangian density
$$
L:J^1Y\op\to^{\wt D}T^*X\op\ot_YVY_\Si\to\op\w^nT^*X
$$
for sections of the composite bundle $Y$.

\section{}

In gravitation theory, we have the composite bundle
$$
LX\to\Si\to X
$$
where $\Si$ is the quotient bundle (\ref{5.15}) and
\[LX_\Si:=LX\to\Si\]
is the L-principal bundle.

Let $P_\Si$ be the $L_s$-principal lift  of $LX_\Si$ such that
\[P_\Si/L_s=\Si, \qquad LX_\Si=r(P_\Si).\]
In particular, there is the imbedding of the $L_s$-lift $P_h$ of $L^hX$
onto the restriction of $P_\Si$ to $h(X)$.

Let us consider the composite spinor bundle (\ref{L1}) where
\[S_\Si= (P_\Si\times V)/L_s\] is
associated with the $L_s$-principal bundle $P_\Si$. It is readily observed
that, given a global section $h$ of $\Si$, the restriction $S_\Si$ to
$h(X)$ is the spinor bundle $S_h$ (\ref{510}) whose sections describe Dirac
fermion fields in the presence of the gravitational field $h$.

Let us provide the principal bundle $LX$ with a holonomic atlas
$\{\psi^T_\xi, U_\xi\}$ and the principal bundles $P_\Si$ and $LX_\Si$
with associated atlases $\{z^s_\e, U_\e\}$ and $\{z_\e=r\circ z^s_\e\}$.
With respect to these atlases, the composite spinor bundle is endowed
with the bundle coordinates $(x^\la,\si_a^\m, \psi^A)$ where $(x^\la,
\si_a^\m)$ are coordinates of the bundle $\Si$ such that
$\si^\m_a$ are the matrix components of the group element
$(\psi^T_\xi\circ z_\e)(\si),$
$\si\in U_\e,\, \pi_{\Si X}(\si)\in U_\xi.$
Given a section $h$ of $\Si$, we have
$$
 (\si^\la_a\circ h)(x)= h^\la_a(x),
$$
where $h^\la_a(x)$ are the tetrad functions (\ref{L6}).

Let us consider the bundle of Minkowski spaces
\[(LX\times M)/L\to\Si\]
associated with the $L$-principal bundle $LX_\Si$. Since $LX_\Si$ is
trivial, it is isomorphic to the pullback $\Si\op\times_X T^*X$
which we denote by the same symbol $T^*X$. Then, one can define the
bundle morphism
\begin{equation}
\g_\Si: T^*X\op\ot_\Si S_\Si= (P_\Si\times (M\ot V))/L_s
\to (P_\Si\times\g(M\ot V))/L_s=S_\Si, \label{L7}
\end{equation}
\[\wh dx^\la=\g_\Si (dx^\la) =\si^\la_a\g^a,\]
over $\Si$. When restricted to $h(X)\subset \Si$,
the morphism (\ref{L7}) comes to the morphism $\g_h$
(\ref{L4}).

We use this morphism in order to construct the total Dirac
operator on the composite spinor bundle $S$ (\ref{L1}).

Let
$$
\wt A=dx^\la\ot (\dr_\la +\wt A^B_\la\dr_B) + d\si^\m_a\ot
(\dr^a_\m+A^B{}^a_\m\dr_B)
$$
be a principal connection on the bundle $S_\Si$ and $\wt D$ the corresponding
differential (\ref{7.10}). We have the
first order differential
operator
\[\cD=\g_\Si\circ\wt D:J^1S\to T^*X\op\ot_SVS_\Si\to VS_\Si,\]
\[\dot\psi^A\circ\cD=\si^\la_a\g^{aA}{}_B(\psi^B_\la-\wt A^B_\la -
A^B{}^a_\m\si^\m_{a\la}),\] on $S$.
One can think of it as being the total Dirac operator since, for every
section $h$, the restriction of $\cD$ to $J^1S_h\subset J^1S$ comes
to the Dirac operator $\cD_h$ (\ref{I13})
relative to the connection
\[A_h=dx^\la\ot[\dr_\la+(\wt A^B_\la+A^B{}^a_\m\dr_\la h^\m_a)\dr_B]
\]
on the bundle $S_h$.

In order to construct the differential $\wt D$ (\ref{7.10}) on $J^1S$
in explicit form, let us consider the principal connection on the bundle
$LX_\Si$ which is given by the local connection form
\ben
&& \wt A = (\wt A^{ab}{}_\m dx^\m+ A^{ab}{}^c_\m d\si^\m_c)\ot I_{ab},
\label{L10}\\
&&\wt A^{ab}{}_\m=\frac12 K^\n{}_{\la\m}\si^\la_c (\eta^{ca}\si^b_\n
-\eta^{cb}\si^a_\n ),\nonumber\\
&&A^{ab}{}^c_\m=\frac12(\eta^{ca}\si^b_\m -\eta^{cb}\si^a_\m),
\label{M4}
\een
where  $K$ is a general linear connection on $TX$ and (\ref{M4})
corresponds to the canonical left-invariant free-curvature connection on
the bundle
\[GL^+(4,\R)\to GL^+(4,\R)/L.\]
 Accordingly, the differential $\wt D$
relative to the connection (\ref{L10}) reads
\begin{equation}
\wt D =dx^\la\ot[\dr_\la -\frac12A^{ab}{}_\m^c (\si^\m_{c\la} +
K^\m{}_{\nu\la}  \si^\nu_c)I_{ab}{}^A{}_B\psi^B\dr_A]. \label{K104}
\end{equation}

Given a section $h$, the connection $\wt A$ (\ref{L10}) is reduced to the
Lorentz connection $\wt K$ (\ref{K102}) on $L^hX$,  and the differential
(\ref{K104}) leads to the covariant derivatives of fermion fields
(\ref{K101}).

We utilize the differential (\ref{K104}) in order to construct a Lagrangian
density of Dirac fermion fields. Their Lagrangian density
is defined on the configuration space $J^1(S\op\oplus_\Si S^+)$ coordinatized
by
$$
( x^\m, \si^\m_a, \psi^A,\psi^+_A, \si^\m_{a\la}, \psi^A_\la, \psi^+_{A\la}).
$$
It reads
\ben
 &&L_\psi=\{\frac{i}2[ \psi^+_A(\g^0\g^\la)^A{}_B( \psi^B_\la -
\frac12A^{ab}{}_\m^c (\si^\m_{c\la} +
K^\m{}_{\nu\la}  \si^\nu_c)I_{ab}{}^B{}_C\psi^C) -\nonumber\\
&& \qquad ( \psi^+_{A\la}- \frac12A^{ab}{}_\m^c (\si^\m_{c\la} +
K^\m{}_{\nu\la}  \si^\nu_c)\psi^+_C
I^+_{ab}{}^C{}_A)(\g^0\g^\la)^A{}_B\psi^B]  -\nonumber\\
&&\qquad m\psi^+_A(\g^0)^A{}_B\psi^B\}\si^{-1}\om \label{230}
\een
where
\[\g^\m=\si^\m_a\g^a, \qquad\si=\det(\si^\m_a),\]
and
$$\psi^+_A(\g^0)^A{}_B\psi^B$$
 is the Lorentz invariant fiber metric in the
bundle $S\op\oplus_\Si S^*$ \cite{cra}.

One can easily verify that
$$
 \frac{\dr\cL_\psi}{\dr K^\m{}_{\nu\la}} +
\frac{\dr\cL_\psi}{\dr K^\m{}_{\la\nu}} =0.
$$
Hence, the Lagrangian density (\ref{230}) depends on the torsion
of the general linear connection $K$ only. In particular, it follows that, if
$K$ is the Levi-Civita connection of a gravitational field $h(x)$,
after the substitution $\si^\nu_c=h^\nu_c(x)$,
the Lagrangian density (\ref{230})
comes to the familiar Lagrangian density of fermion fields in the Einstein's
gravitation theory.

\bigskip

\centerline{\bf Acknowledgement}
\medskip

The authors would like to thank Prof. L.Mangiarotti and Prof. Y.Obukhov for
valuable discussions.


\begin{thebibliography}{ederf}

\bibitem{ar} A.Aringazin and A.Mikhailov, Matter fields in spacetime with
vector non-metricity, {\it Clas. Quant. Grav.} {\bf 8}, 1685 (1991).

\bibitem{cra} J.Crawford, Clifford algebra: Notes on the spinor metric and
Lorentz, Poincar\'e and conformal groups, {\it J. Math. Phys.} {\bf 32}, 576
(1991).

\bibitem{fer} M.Ferraris and J.Kijowski, On the equivalence of the
relativistic theories of gravitation, {\it GRG} {\bf 14}, 165 (1982).

\bibitem{cam2} G.Giachetta and G.Sardanashvily, Stress-energy-momentum
of affine-metric gravity. Generalized Komar superpotential, E-print:
gr-qc/9511008.

\bibitem{heh} F.Hehl, J.McCrea, E.Mielke and Y.Ne'eman: Metric-affine
gauge theory of gravity, {\it Phys. Rep.} {\bf 258}, 1 (1995).

\bibitem{kob} S.Kobayashi and K.Nomizu, {\it Foundations of Differential
Geometry, Vol.1}  (John Wiley, N.Y. - Singapore, 1963).

\bibitem{mag} G.Magnano, Are there metric theories of gravity other than
General Relativity, E-print: gr-qc/9511027.

\bibitem{pon} V.Ponomarev and Yu.Obukhov, Generalized Einstein-Maxwell
theory, {\it GRG} {\bf 14}, 309 (1982).

\bibitem{sard9}
G.Sardanashvily, On the geometry of spontaneous symmetry breaking,
{\it J. Math. Phys.} {\bf 33}, 1546 (1992).

\bibitem{sard92}
G.Sardanashvily and O. Zakharov, {\it Gauge Gravitation Theory},
 (World Scientific, Singapore, 1992).

\bibitem{sard} G.Sardanashvily, {\it Gauge Theory in Jet Manifolds}
(Hadronic Press, Palm Harbor, 1993).

\bibitem{sard0}
G.Sardanashvily, Five lectures on the jet manifold methods in field
theory, E-print: hep-th/9411089.

\bibitem{sard1}
G.Sardanashvily, Composite spinor bundles in gravitation theory,
E-print: gr-qc/9502022.

\bibitem{sau} D.Saunders, {\it The Geometry of Jet Bundles}
(Cambridge Univ. Press, Cambridge, 1989).

\bibitem{art} G.Thompson, Non-uniqueness of metrics compatible with a
symmetric connection, {\it Class. Quant. Grav.}, {\bf 10}, 2035 (1993).

\bibitem{tuc} R.Tucker and C.Wang, Black holes with Weyl charge and
non-Riemannian waves, {\it Class. Quant. Grav.} {\bf 12}, 2587 (1995).
\end{thebibliography}
\end{document}